\documentclass[10pt]{article}

\usepackage{geometry}
\usepackage{hyperref}

\usepackage{cite}

\usepackage[cmex10]{amsmath}
\usepackage{amssymb}
\usepackage{amsthm}

\usepackage{array}

\usepackage{float}
\usepackage{algpseudocode}
\floatstyle{ruled}
\newfloat{algorithm}{ht}{loa}
\floatname{algorithm}{Algorithm}
\newcommand{\Input}{\textbf{Input:\ }}
\newcommand{\Output}{\textbf{Output:\ }}

\usepackage{rotating}

\usepackage{functan}
\Macro{L2}{\mathrm{L}^2(\Omega)}
\Macro{l0}{0}
\Macro{l1}{1}
\Macro{l2}{2}
\Macro{lp}{p}
\Macro{l2l2}{\ell_2\rightarrow\ell_2}
\Macro{l2w}{\ell^2(\vec{w})}
\Macro{lo}{\infty}
\Macro{G}{\mathscr{G}}
\Macro{H}{\mathscr{H}}
\Macro{K}{\mathscr{G}}
\Macro{K2}{K\rightarrow 2}
\Macro{S2}{S\rightarrow 2}

\newcommand{\abs}[1]{\left\vert #1 \right\vert}
\renewcommand{\vec}[1]{\mathbf{#1}}

\usepackage{mathrsfs} 
\newcommand{\listspacing}{\setlength{\itemsep}{-1mm}}

\DeclareMathOperator*{\argmin}{arg\ min}

\DeclareMathOperator{\supp}{supp}

\renewcommand{\it}[2]{#1^{(#2)}}

\renewcommand{\H}{\mathscr{H}}
\newcommand{\Ht}{\mathbb{H}}

\newcommand{\PP}{\mathfrak{P}}

\newcommand{\R}{\mathbb{R}}

\newcommand{\Rmn}{\mathbb{R}^{m\times n}}
\newcommand{\Z}{\mathbb{Z}}

\renewcommand{\a}{\vec{a}}
\newcommand{\bsh}{\beta_{s+\hat{s}}}

\newcommand{\ds}{\delta_s}
\newcommand{\dsh}{\delta_{s+\hat{s}}}
\newcommand{\dshh}{\delta_{s+2\hat{s}}}
\newcommand{\e}{\vec{e}}
\newcommand{\etol}{\varepsilon_{{tol}}}

\newcommand{\gk}{\it{\gamma}{k}}

\renewcommand{\r}{\vec{r}}
\newcommand{\s}{\vec{s}}
\newcommand{\sh}{\hat{s}}
\renewcommand{\u}{\vec{u}}
\renewcommand{\v}{\vec{v}}

\newcommand{\x}{\vec{x}}

\newcommand{\xh}{\hat{\x}}
\newcommand{\y}{\vec{y}}
\newcommand{\z}{\vec{z}}

\newcommand{\define}{\triangleq}
\newcommand{\trans}{\mathsf{T}}

\usepackage{extarrows}
\newtheorem{thm}{Theorem}
\newtheorem{cor}[thm]{Corollary}
\newtheorem{lem}[thm]{Lemma}
\newtheorem{prop}[thm]{Proposition}

\renewcommand{\ge}{\geqslant}
\renewcommand{\le}{\leqslant}

\usepackage{setspace}

\usepackage{url}
\begin{document}
\title{Modified Frame Reconstruction Algorithm\\ for Compressive Sensing}
\author{Graeme~Pope}

\maketitle

\begin{abstract}
Compressive sensing is a technique to sample signals well below the
Nyquist rate using linear measurement operators.  In this paper we
present an algorithm for signal reconstruction given such a set of
measurements.  This algorithm generalises and extends previous
iterative hard thresholding algorithms and we give sufficient
conditions for successful reconstruction of the original data signal.
In addition we show that by underestimating the sparsity of the data
signal we can increase the success rate of the algorithm.

We also present a number of modifications to this algorithm:\ the
incorporation of a least squares step, polynomial acceleration and an
adaptive method for choosing the step-length.  These modified
algorithms converge to the correct solution under similar conditions
to the original un-modified algorithm.  Empirical evidence show that
these modifications dramatically increase both the success
rate and the rate of convergence, and can outperform other algorithms
previously used for signal reconstruction in compressive sensing.
\end{abstract}

\section{Introduction}
{C}{ompressive} sensing is a radical new way of sampling signals at a sub-Nyquist rate. The Shannon/Nyquist sampling theorem tells us that an analogue signal can be reconstructed perfectly from its samples, if it was sampled at a rate at least twice the highest frequency present in the signal, known as the Nyquist rate \cite{Nyquist1928,Shannon1949}.  For many signals, such as audio or images, the Nyquist rate can be very high.  This may result in acquiring a very large number of samples, which must be compressed in order to store or transmit them, as well as placing a high requirement on the equipment needed to sample the signal. \emph{Compressive Sensing} (or compressed sensing or CS) is a recently introduced method  that can reduce the number of measurements required, in some ways it can be regarded as automatically compressing the signal.  Compressive sensing is a technique that enables us to fully reconstruct particular classes of signals if the original signal, is sampled at a rate well \emph{below} the Nyquist rate.

In particular, compressive sensing works with sparse signals.  In many applications the signal of interest is primarily zero in some known fixed basis, that is, in this representation the data contained within the signal is sparse.  Traditional measurement techniques which consist of sampling at all the possible data points heavily over-sample the signal.  Consider the scenario where we randomly draw samples from a sparse signal, then the probability of sampling at an ``interesting'' data point is equal to the sparsity fraction.  Compressive sensing gets around this by using a small number of linear sampling operators that samples the signal across all data points simultaneously.  This gives rise to the name compressive sensing as it is a combination of sampling and compression.

More formally, let $\x\in\R^n$ be an $s$-sparse vector in the basis $\Psi$, then we can write $\x=\Psi\s$ for some vector $\s\in\R^n$ such that only $s\le n$ components of $\s$ are non-zero.  Assume then that our $m$ linear sampling operators are given by the rows of the matrix $\Phi\in\Rmn$.  The problem of compressive sensing is then to find $\x$, or equivalently, $\s$ given the measurements $\y=\Phi\x$ and the matrices $\Phi$ and $\Psi$, under the assumption that $\s$ is sparse.  If $m<n$ then the problem is under-determined and there is no unique solution, so we commonly say that we are interested in the most sparse solution to this equation, i.e. the minimiser to
\begin{equation*}
  \min_{\hat{s}\in\R^n} \norm{l0}{\hat{s}} \quad \mbox{subject to}\quad \y=\Phi\Psi\hat{s}.
\end{equation*}
To solve this problem in general is known to be NP-hard \cite{Davis1994,Natarajan1995}.

One of the original breakthroughs in compressive sensing  was to show that linear programming methods can be used to efficiently reconstruct the data signal with high accuracy \cite{Candes2005,Candes2005a,Donoho2006}.   Since then many alternative methods have been proposed as a faster or more successful alternative to these linear programming algorithms.  One approach is to use matching pursuit techniques, originally proposed in \cite{Mallat1993}, variations have been proposed such as OMP or orthogonal matching pursuit \cite{Tropp2007}, Stagewise orthogonal matching pursuit (StOMP) \cite{Donoho2007}, Compressive sampling matching pursuit (CoSaMP) \cite{Needell2008} and gradient pursuit algorithms \cite{Blumensath2008d,Blumensath2008a,Blumensath2008b,Blumensath2008c}.  Also proposed has been a suite of thresholding based algorithms, either hard thresholding (IHT) \cite{Blumensath2008,Blumensath2009} or soft thresholding \cite{Daubechies2004,Daubechies2007}.  What we propose is a combination of some of these techniques, combining hard thresholding with matching pursuit methods.  We will also show how we can use polynomial acceleration techniques to increase the rate of convergence.

In addition to this, work has also been done on model based compressive sensing in \cite{Baraniuk2008}, which can be applied to many of the algorithms above.  Most of the aforementioned algorithms, in particular CoSaMP and IHT make use of a pruning step which takes a solution and forces it to be sparse, by removing all but the $\sh$-largest (in magnitude) components, which is the best $\sh$ sparse approximation under any $\ell_p$ norm for $1\le p < \infty$.  Model based compressive sensing proposes using a model based algorithm to perform this, that is to choose the sparse signal that is not necessarily closest under an $\ell_p$ distance, but that best fits the signal model.  Such a modification would also be applicable to our algorithm.

We will refer to our algorithm as the ``Modified Frame Reconstruction'' or MFR algorithm.

\subsection{Paper Overview}
In section \ref{sec:background} of this paper we will give an overview of the fundamental concepts in compressive sensing.  We will also briefly discuss frames, the frame reconstruction algorithm and polynomial acceleration using Chebyshev polynomials.  In section \ref{sec:mfr} we will present our own algorithm and theoretical results regarding its performance.  This work is similar to that in \cite{Blumensath2008,Blumensath2009}, but we analyse the algorithm in a different manner and in a more general setting. Later in section \ref{sec:modifications} we present a number of extensions to this algorithm that increase both the rate of convergence and the probability of successfully estimating the original input signal.  We show that by incorporating a least squares step, polynomial acceleration and a variable step-length, that we can significantly increase the performance of our algorithm.  In section \ref{sec:results} we perform simulations of our algorithm demonstrating that the theoretical results translate into a real world performance advantage.  Section \ref{sec:comparison} contains a comparison of our algorithm to various other existing methods.

\subsection{Notation}
In this paper  bold-faced symbols represent vectors, such as $\x\in\R^n$, with components $x_1,\ldots,x_n$. For an index set $\Gamma\subset\{1,2,\ldots,n\}$ we take $\x_{\Gamma}\in\R^n$ to be the vector that agrees with $\x$ on all the indices $i\in\Gamma$ and is 0 elsewhere.  For a matrix $\Phi\in\Rmn$ we take $\Phi^\trans,\Phi^*$ and $\Phi^\dagger$ to be the transpose, complex-conjugate transpose and Moore-Penrose pseudo-inverse respectively.  Hence $\Phi^\dagger \y$ is the solution to the minimisation problem $\arg \min_\x \norm{l2}{\Phi\x - \y}$.
For a matrix $\Phi$ we take $\Phi_{\Gamma}$ to be the submatrix formed by taking the columns of $\Phi$ indicated by the set $\Gamma$.  When we write $\Phi_\Gamma^\trans$ we apply the column selection first, then the transpose, e.g.\ $\Phi_\Gamma^\trans \define \left(\Phi_\Gamma\right)^\trans$.  We take $\lambda_{\min}(\Phi),\lambda_{\max}(\Phi)$ to be the lower and upper eigenvalues of the matrix $\Phi$ and similarly $\sigma_{\max}(\Phi),\sigma_{\min}(\Phi)$ are the upper and lower singular values of $\Phi$.

Recall that the $\ell_p$ distance, for $1\le p <\infty$, between two vectors $\x,\y\in\R^n$ is defined to be $d(\x,\y)\define \left(\sum_{i=1}^n \abs{x_i-y_i}^p\right)^{1/p}$ giving rise to the $\ell_p$ norm $\norm{lp}{\x} \define \left(\sum_{i=1}^n \abs{x_i}^p \right)^{1/p}$.  The $\ell_0$ pseudo-norm is given by $\norm{l0}{\x} \define \abs{\{x_i \neq 0\}}$, i.e.\ the number of non-zero components of $\x$.  We say that a vector $\x$ is $s$-sparse if $\norm{l0}{\x} = s$, that is, precisely $s$ components of $\x$ are non-zero.  The $i$-th stage of an iterative algorithm producing the vector $\x$ is denoted by $\it{\x}{i}$.  

We will write $\Ht_\tau:\R^n \rightarrow \R^n$ to be the function that returns $\xh$, a best $\tau$-sparse vector approximation to the input $\x$ under any $\ell_p$-norm, $1\le p < \infty$, i.e.\ a $\tau$-sparse vector such that $\norm{lp}{\x-\xh}$ is minimal.  Note that this has the same solution for any $1\le p<\infty$: the vector $\xh$ consisting of the $\tau$-largest (in magnitude) components of $\x$.  If this vector is not unique, we can decide whether to break the tie randomly or deterministically.  In our implementation we chose to break the tie lexicographically.  We also write $\supp(\x)$ to denote the support of the vector $\x$, that is, $\supp(\x)\define \{x_i\colon x_i\neq 0 \}$.

\section{Background} \label{sec:background}

\subsection{Compressive Sensing}
Let $\x\in\R^n$ be a signal and let $\Phi\in\Rmn$ be a measurement matrix.  We call the vector $\y\in\R^m$ the vector of observations if $\y = \Phi \x + \e$ where $\e\in\R^m$ is a vector of noise or errors, possibly equal to $\vec{0}$.  The task is to recover $\x$ given only the observations $\y$ and the matrix $\Phi$.  Clearly if $\e=\vec{0}$, $m=n$ and the columns of $\Phi$ are linearly independent, then $\Phi$ is invertible and $\x$ can be recovered exactly.  However compressive sensing asks (and answers) the question, how well can we do if $m \ll n$?

First, let us define the \emph{Restricted Isometry Property} (RIP) condition of order $s$ for a matrix $\Phi\in\Rmn$ as per \cite{Candes2006a}.  Let $\delta_s\ge 0$ be the smallest value such that
\begin{equation}
	1-\delta_s\le \frac{ \norm{l2}{\Phi\x}^2}{\norm{l2}{\x}^2} \le 1+\delta_s
\end{equation}
for all $s$-sparse vectors.  We call $\delta_s$ the \emph{restricted isometry constant of order $s$}.  If $\delta_s<1$ we say that $\Phi$ satisfies the RIP of order $s$.  It is easy to see that we have
\begin{equation*}
	\lambda_{\min}\left(\Phi^\trans\Phi\right) \le 1-\ds \le 1+\ds \le \lambda_{\max}\left(\Phi^\trans\Phi\right),
\end{equation*}
for any $s$.  For a given matrix $\Phi$ the restricted isometry constant $\delta_s$ can be calculated by
\begin{equation*}
	\ds = \max_{\Gamma\colon \abs{\Gamma} = s} \left\{ 1-\lambda_{\min}\left(\Phi_\Gamma^\trans\Phi_\Gamma\right),\ \lambda_{\max}\left(\Phi_\Gamma^\trans\Phi_\Gamma\right) - 1\right\}.
\end{equation*}
This is proved in Lemma \ref{lem:eig_values_RIP}.

In order to give some sense to the problem of finding a vector $\x$, given the measurements $\y$ so that $\y=\Phi\x$, we make the assumption that $\x$ is sparse.  We can then cast this question as a minimisation problem
\begin{equation*}
	\arg \min_{\xh\in\R^n} \norm{l0}{\xh} \text{ subject to } \y=\Phi\xh.
\end{equation*}
The following lemma from \cite{Candes2005a} gives conditions under which the minimiser to the $\ell_0$ minimisation problem above is the same as the solution to $\y=\Phi\x$.
\begin{lem}[Lemma 1.2 of \cite{Candes2005a}] \label{lem:existence}
Let $\Phi\in\Rmn$ be a matrix with RIP constant $\delta_{2s}< 1$ and let $\Gamma$ be an index set with $\abs{\Gamma}\le s$.  Let $\x\in\R^n$ be a vector with support $\Gamma$ and set $\y=\Phi\x$.  Then $\x$ is the unique solution to 
\begin{equation*}
	\arg\min_{\xh\in\R^n} \norm{l0}{\xh} \qquad \text{subject to} \qquad \y=\Phi\xh,
\end{equation*}
and hence $\x$ can be reconstructed exactly from  $\y$.
\end{lem}

Lemma \ref{lem:existence} only proves the existence of a solution but does not say anything about how to find it.  

One question is, what matrices $\Phi$ satisfy the RIP of order $s$?  In general, it is not possible to design a matrix $\Phi$ that satisfies the RIP of a particular order.  It has been shown that certain classes of random matrices will obey the RIP with very high probability.  For example, let the columns of $\Phi$ be sampled uniformly at random from the unit sphere, or let each entry $\Phi_{ij}$ be sampled independently from the Gaussian distribution with mean 0 and variance $\frac 1 m$.  Then  $\Phi$ obeys the RIP of order $s$ with probability at least $1-\mathcal{O}(e^{-m})$ provided that $m \ge \mathrm{C} s \log(n/s)$ for some constant $\mathrm{C}$.  As a rule of thumb, for a Gaussian measurement matrix it suffices to take $m\approx 2s\log n$ so that $\ell_1$ minimisation will work. \cite{Candes2006c,Baraniuk2007b,Donoho2006a}

\subsection{Frames}
Frames are a generalisation of bases for Hilbert spaces and have been heavily used in wavelet decompositions \cite{Daubechies1990,Heil1989}.  Let $\H$ be an $m$-dimensional Hilbert space and let $\Phi = \{\varphi_i \in\H \colon 1\le i \le n\}$ be a set of $n$ elements.  Then $\Phi$ forms a frame for $\H$ if there exist constants $0<A\le B < \infty$ such that 
\begin{equation} \label{eq:frame}
	A \norm{H}{x}^2 \le \sum_{i=1}^n \abs{\scalprod{H}{x}{\varphi_i}}^2 \le B \norm{H}{x}^2,
\end{equation}
for all $x\in\H$ and where $\norm{H}{\cdot}$ and $\scalprod{H}{\cdot}{\cdot}$ are the norm and scalar-product defined on the Hilbert space. The function $S:\H\rightarrow\H$ given by
\begin{equation*}
	Sx \define \sum_{j} \scalprod{H}{x}{\varphi_j}\varphi_j,
\end{equation*}
is called the \emph{frame operator}.  Then given the vector $\y$ of observations where $y_i = \scalprod{H}{x}{\varphi_i}$ we can reconstruct the original element $x\in\H$ via the iterative algorithm 
\begin{equation} \label{eq:frame_alg}
	\it{x}{i+1} = \it{x}{i} + \frac{2}{A+B}S\left(x - \it{x}{i}\right),
\end{equation}
where $A$ and $B$ are the frame bounds in \eqref{eq:frame} \cite{Heil1989}.  Note that although $x$ is unknown, we know $Sx$ from the measurement vector $\y$ and hence this algorithm can be implemented since $S$ is linear.  This algorithm will converge from any starting element $x\in\H$.  In particular if $\it{x}{0} \equiv 0$, then the error satisfies
\begin{equation*}
	\norm{H}{x-\it{x}{k}} \le \left(\frac{B-A}{B+A}\right)^k \norm{H}{x}.
\end{equation*}
Faster algorithms also exist, which can be found, for example, in \cite{Grochenig1993}.  

It was shown in \cite{Grochenig1993} that using any positive value $\gamma\le 2/\left({B+A}\right)$ in place of $2/\left({B+A}\right) $ will also result in correct reconstruction, but at a possibly slower rate of convergence.

This algorithm is identical to some of those used in iteratively solving least squares problems, such as \emph{Richardson's first order method} \cite[pp276-280]{Bjoerck1996}.

\subsection{Polynomial Acceleration}\label{sec:accelerated}
Iterative algorithms can often be sped up by considering the output of all the previous iterations, not just the very last iteration.  \emph{Semi-iterative} methods such as \emph{polynomial acceleration} \cite{Golub1961,Golub1961a} and \emph{Richardson's second order method} \cite[pp280-282]{Bjoerck1996} are two ways of doing this.  The idea behind these two methods is to use the solution from previous iterations for example, if we use the update
\begin{equation*}
  \it{\x}{k} = \sum_{i=0}^{k-1} c_{k,i} \it{\x}{i},\quad  \sum_{i=0}^{k-1} c_{k,i} = 1.
\end{equation*}
Define the polynomial $P_k(t)$ by
\begin{equation*}
  P_k(t) \define \sum_{i=0}^k c_{k,i} t^i, \quad P_k(1)=1,
\end{equation*}
and it follows that the error equation is
\begin{equation*}
  \it{\x}{k}-\x = P_k(B) \left(\it{\x}{0}-\x\right),
\end{equation*}
where $P_k(B)$ is a polynomial in the matrix $B$, hence the name \emph{polynomial acceleration}.  The convergence rate is bounded above by
\begin{equation*}
  \max_{\lambda_{\min}(B) \le t \le \lambda_{\max}(B)} \abs{P_k(t)},
\end{equation*}
and so we wish to find the minimiser to
\begin{equation*}
  \min_{P_k:\deg P_k = k} \max_{t\in[0,1]} \abs{P_k(t)}.
\end{equation*}
The minimising class of polynomials for this term is the Chebyshev polynomials of the first kind \cite{Young1971}, defined by the recurrence relationship
\begin{equation*}
  T_{n+1}(x) = 2xT_n(x)-T_{n-1}(x),
\end{equation*}
 with $T_0(x)=1$ and $T_1(x)=x$.  It is shown in \cite{Golub1961} that we do not need to use all the previous iterates, as the polynomials are generated recursively it suffices to use
\begin{equation*}
  \it{\x}{1} = \it{\x}{0} + \gamma A^\trans \left(\y-A\it{\x}{0}\right)
\end{equation*}
and
\begin{equation*}
  \it{\x}{k+1} = 
    \it{\x}{k-1}+\it{\omega}{k+1}\left(\gamma A^\trans\left(\y-A\it{\x}{k}\right) +\it{\x}{k} - \it{\x}{k-1}\right),
\end{equation*}
where
\begin{equation*}
  \it{\omega}{k+1} = \frac{1}{1-\it{\omega}{k}\frac{\mu^2}{4}}, \quad \it{\omega}{1}=1, \quad \mu = \frac{b-a}{b+a}
\end{equation*}
and where $a,b$ are the minimum and maximum eigenvalues of $A^\trans A$.  This particular method is known as the \emph{Chebyshev semi-iterative method}.  The second order Richardson method differs only slightly in that we take 
\begin{equation*}
    \omega = \it{\omega}{k} = \frac{2}{1+\sqrt{1-\mu^2}}.
\end{equation*}
In fact it can be shown that in the Chebyshev method we have $\it{\omega}{k}\rightarrow \omega$ as $k\rightarrow \infty$.  These algorithms typically converge an order of magnitude faster than without the acceleration \cite{Bjoerck1996}.  In \cite{Grochenig1993} it is also shown that this technique leads to significantly faster convergence of the frame reconstruction algorithm.
  
\section{Reconstruction Algorithm} \label{sec:mfr}
\subsection{MFR -- Modified Frame Reconstruction Algorithm}
Our approach is based on the frame reconstruction algorithm in \eqref{eq:frame_alg}.  The key observation for MFR is the fact that we can still perform this iterative step even if the matrix $\Psi$ no longer forms a frame for the space $\R^n$.  And that if the algorithm converges, it can still converge to a solution $\xh$ of the now under-determined matrix equation $\y=\Phi\x$, which is exactly the problem of compressive sensing.

We give the plain version of the MFR algorithm (Algorithm \ref{alg:mfr_alg}) and then show several modifications   which increase both the rate of convergence and the success rate in finding the original sparse vector.  We will also show theoretical bounds for the convergence and give sufficient conditions for convergence to occur.

The algorithm consists of two parts, an update and a thresholding step.
\begin{enumerate} 
\item \textbf{Update: } Similar to the frame algorithm we perform an update
\begin{equation}
	\it{\a}{k+1} = \it{\x}{k} + \gamma \Phi^\trans \left(\y - \Phi\it{\x}{k}\right),
\end{equation}
where $\y$ is the vector of measurements, $\Phi$ is the measurement matrix and $\gamma$ is a control parameter which we often refer to as the \emph{step-length}. 
\item \textbf{Thresholding: } The second part of the algorithm is the thresholding procedure where we generate the next ``solution''
\begin{equation}
	\it{\x}{k+1} = \Ht_{\sh}\left[\it{\a}{k+1}\right].
\end{equation}
Here we simply threshold the output of the frame step producing an $\sh$-sparse approximation. 
\end{enumerate}
 Recall that $\Ht_{\sh}(\z)$ produces the best $\sh$-sparse approximation to the input $\z$ under any $\ell_p$ norm for $1\le p<\infty$.  These two steps are repeated until ``convergence'' occurs, that is, the change from one iteration to the next is sufficiently small.

\begin{algorithm}[!t] 
   \begin{flushleft}
     \caption{Modified Frame Reconstruction Algorithm} 
     \label{alg:mfr_alg}
     \Input 
     \begin{itemize} 
     \item The measurement matrix $\Phi$.
     \item Observation vector $\y$.
     \item Estimate of sparsity $\hat{s}$ of the vector $\x$.
     \item Step size $\gamma$.
     \item Tolerance parameter $\etol$.
     \end{itemize}
     
\Output
\begin{itemize} 
	\item A vector $\xh\in\R^n$ that is $\sh$-sparse.
\end{itemize}

\begin{algorithmic}[1]

	\State $\it{\x}{0}\leftarrow \vec{0}$
        \State $k\leftarrow 1$
	\While{ $ \norm{l2}{\it{\x}{k} - \it{\x}{k-1} } \ge \etol$}
		\State $\it{\x}{k+1} \leftarrow \Ht_{\hat{s}}\left[ \it{\x}{k} + \gamma \Phi^\trans\left(\y - \Phi\it{\x}{k}\right)\right]$ 
                \State $k\leftarrow k+1$
	\EndWhile
	\State \Return $\it{\x}{k}$
\end{algorithmic}
\end{flushleft}
\end{algorithm}

\subsection{Analysis of Performance}
We now state the properties of the MFR algorithm in Theorem \ref{thm:main} and Proposition \ref{prop:main}.  Proposition \ref{prop:main} gives sufficient convergence conditions in the scenario where we measure a sparse signal with noise, i.e.\ in the model $\y=\Phi\x+\e$ where $\Phi\in\Rmn$ is the measurement matrix, $\x\in\R^n$ is the $s$-sparse data, $\e\in\R^m$ is noise and $\y\in\R^m$ is the observed signal.  We will see in Theorem \ref{thm:main} that the MFR algorithm converges even when the signal is not sparse.  The proof of these results uses the same techniques as the proof in \cite{Blumensath2008} although we choose to use the RIP condition rather than the modified RIP condition.
\begin{thm} \label{thm:main}
Fix $\sh=s$ where $s$ is the sparsity of the desired solution.  Then given measurements $\y=\Phi\x+\e$ where $\Phi\in\Rmn$ has the RIP such that either condition (a), (b) or (c) is satisfied
\begin{subequations}
\begin{align*}
	&\mbox{(a) } \gamma \ge \frac{1}{\delta_{3s}-\delta_{2s}+1}, \mbox{ and }\gamma \delta_{3s} \le \frac{1}{\sqrt{32}},  \mbox{ or} \\
	&\mbox{(b) } \gamma < \frac{1}{\delta_{3s}-\delta_{2s}+1}, \mbox{ and }\gamma(1-\delta_{2s}) \ge 1- \frac{1}{\sqrt{32}}, \mbox{ or} \\
	&\mbox{(c) }  \frac{3}{4(1-\delta_{2s})} < \gamma < \frac{1}{1-\delta_{2s}} \mbox{ and } \delta_{2s} < 1, 
\end{align*}
\end{subequations}
then  Algorithm \ref{alg:mfr_alg} will recover an approximation $\it{\x}{k}$ satisfying
\begin{align}
	\norm{l2}{\it{\x}{k} - \x^{s}} &\le 2^{-k} \norm{l2}{\x^{s}}+ 4\gamma\sqrt{1+\delta_{2s}}\norm{l2}{\e} + \ldots \nonumber \\
	&\quad 4\gamma\big(1+\delta_{2s}\big)\left(\norm{l2}{\x-\x^{2s}} + \frac{1}{\sqrt{2s}}\norm{l1}{\x-\x^{2s}}\right),
\end{align}
where $\x^{2s}$ is the best $2s$-sparse approximation to $\x$.
\end{thm}

\begin{prop} \label{prop:main}
Under the conditions of Theorem \ref{thm:main} and given measurements $\y=\Phi\x+\e$ where $\x$ is $s$-sparse for $s\le \sh$ and $\Phi\in\Rmn$ has the RIP such that either condition (a), (b) or  (c) is satisfied
\begin{align*}
	&\mbox{(a) } \gamma \ge \frac{1}{\dshh-\dsh+1}, \mbox{ and }\gamma \dshh \le \frac{1}{\sqrt{32}}, \mbox{ or} \\
	&\mbox{(b) } \gamma < \frac{1}{\dshh-\dsh+1}, \mbox{ and } \gamma(1-\dsh) \ge 1- \frac{1}{\sqrt{32}},\mbox{ or} \\
	&\mbox{(c) }  \frac{3}{4(1-\dsh)} < \gamma < \frac{1}{1-\dsh}\mbox{ and } \dsh < 1,
\end{align*}
then Algorithm \ref{alg:mfr_alg} will recover an approximation $\it{\x}{k}$ satisfying
\begin{equation}
	\norm{l2}{\x-\it{\x}{k}} \le 2^{-k} \norm{l2}{\x} + 4\gamma\sqrt{1+\dsh} \norm{l2}{\e}.
\end{equation}
\end{prop}

\begin{proof}
Conditions (a) and (b) come from Lemma \ref{lem:main} and condition (c) follows by setting $\alpha=\frac{1}{2}$ in Lemma \ref{lem:theoretical_max}.
\end{proof}

We will first prove Lemmas \ref{lem:main} and \ref{lem:theoretical_max}, hence proving Proposition \ref{prop:main} and then use this to prove the main theorem.

\begin{lem} \label{lem:main}  
Under the conditions of Theorem \ref{thm:main} and given measurements $\y=\Phi\x+\e$ where $\x$ is $s$-sparse for $s\le \sh$ and $\Phi\in\Rmn$ has the RIP such that either condition (a) or condition (b) is satisfied
\begin{subequations}
\begin{align*}
	&\mbox{(a)} &\gamma \ge \frac{1}{\dshh-\dsh+1}, &\mbox{ and } \gamma \dshh \le \frac{1}{\sqrt{32}}, \quad \mbox{or} \\
	&\mbox{(b)} &\gamma < \frac{1}{\dshh-\dsh+1}, &\mbox{ and } \gamma(1-\dsh) \ge 1- \frac{1}{\sqrt{32}}, 
\end{align*}
\end{subequations}
then Algorithm \ref{alg:mfr_alg} will recover an approximation $\it{\x}{k}$ satisfying
\begin{equation}
	\norm{l2}{\x-\it{\x}{k}} \le 2^{-k} \norm{l2}{\x} + 4\gamma\sqrt{1+\dsh} \norm{l2}{\e}.
\end{equation}
\end{lem}

To prove Lemma \ref{lem:main}, we need the following lemmas.
\begin{lem} \label{lem:eig_values_RIP}
Suppose that $\Phi\in\Rmn$ obeys the restricted isometry property of order $s$ with value $\delta_s$.  Then for any set of indices $\Gamma$ such that $\abs{\Gamma}\le s$ the singular values of $\Phi_\Gamma$ lie in the range $[\sqrt{1-\delta_s},\sqrt{1+\delta_s}]$.  Furthermore the eigenvalues of $\Phi^\trans_\Gamma \Phi_\Gamma - \vec{I}$ lie in the interval $[-\delta_s,\delta_s]$.
\end{lem}
\begin{proof}
Recall that the RIP says that for all vectors $\x'\in\R^t$ for $t\le s$
\begin{equation*}
	1-\delta_s \le \frac{\norm{l2}{\Phi_\Gamma \x'}^2}{\norm{l2}{\x'}^2} \le 1+\delta_s.
\end{equation*}
Furthermore we know that the Rayleigh quotient for a matrix $A$ is bounded by the minimum and maximum eigenvalues of $A$, i.e.\ $\lambda_{\min}(A) \le \rho_A(\x) \le \lambda_{\max}(A)$ for all $\x$.  Set $A = \Phi^\trans_\Gamma \Phi_\Gamma$, then the eigenvalues of $A$ are the square of the singular values of $\Phi_\Gamma$.  As the eigenvalues for $A$ are bounded by $1\pm \delta_s$ we have the corresponding bound for the singular values of $\Phi_\Gamma$, namely $\sigma(\Phi_\Gamma) \in [\sqrt{1-\delta_s},\sqrt{1+\delta_s}]$.  It follows then that the eigenvalues of  $\Phi^\trans_\Gamma \Phi_\Gamma - \vec{I}$ lie in the interval $[-\delta_s,\delta_s]$.
\end{proof}

\begin{lem} \label{lem:blu_1}
For all index sets $\Gamma$ and all measurement matrices $\Phi$ for which the RIP holds with $s=\abs{\Gamma}$
\begin{equation}
	\norm{l2}{\left(\vec{I} - \Phi_{\Gamma}^\trans \Phi_{\Gamma}\right)\x_\Gamma} \le \delta_s \norm{l2}{\x_{\Gamma}}.
\end{equation}
and furthermore 
\begin{equation}
	\norm{l2}{\left(\vec{I} - \gamma \Phi_{\Gamma}^\trans \Phi_{\Gamma}\right)\x_\Gamma} \le \left[1-\gamma(1-\delta_s )\right]\norm{l2}{\x_{\Gamma}}.
\end{equation}
\end{lem}
\begin{proof}
The RIP guarantees us that the eigenvalues of $ \Phi_{\Gamma}^\trans \Phi_{\Gamma}$ lie in the range $1-\delta_s$ to $1+\ds$.  Hence the matrix $\vec{I} - \Phi_{\Gamma}^\trans \Phi_{\Gamma}$ has eigenvalues in $[-\ds,\ds]$.  For the second result we proceed similarly.  Clearly $\gamma \Phi_{\Gamma}^\trans \Phi_{\Gamma}$ has eigenvalues in the range $[\gamma(1-\ds),\gamma(1+\ds)]$, hence the maximum eigenvalue of $\vec{I} - \gamma \Phi_{\Gamma}^\trans \Phi_{\Gamma}$ is less than or equal to $1-\gamma(1-\ds)$.  

Then for each fixed $\Gamma$ we have
\begin{subequations}
\begin{align*}
	\norm{l2}{\left(\vec{I} - \gamma \Phi_{\Gamma}^\trans \Phi_{\Gamma}\right)\x_\Gamma} & \le \norm{l2}{\left(\vec{I} - \gamma \Phi_{\Gamma}^\trans \Phi_{\Gamma}\right)}\cdot \norm{l2}{\x_\Gamma} \\
	&\le \left(1-\gamma(1-\ds)\right)\norm{l2}{\x_\Gamma},
\end{align*}
\end{subequations}
and the first equation follows by setting $\gamma=1$.
\end{proof}

\begin{lem}[Based on Lemma 2 of \cite{Blumensath2008}] \label{lem:eig_values}
Let $\Gamma$ and $\Lambda$ be two disjoint index sets for the matrix $\Phi$.  Then for all $\Phi$ for which the RIP holds with $s=\abs{\Gamma\cup\Lambda}$
\begin{equation}
	\norm{l2}{\Phi_\Gamma^\trans \Phi_\Lambda \x_\Lambda} \le \ds \norm{l2}{\x_\Lambda}.
\end{equation}
\end{lem}

\begin{proof}
Set $\Omega = \Gamma\cup\Lambda$.  Since $\Gamma$ and $\Lambda$ are disjoint, the matrix $\left(-\Phi_\Gamma^\trans \Phi_\Lambda\right)$ is a submatrix of the matrix $\vec{I} - \Phi^\trans_\Omega \Phi_\Omega$.  Since the largest singular value of a submatrix is bounded above by the largest singular value of the full matrix, we have
\begin{equation*}
	\norm{l2}{\Phi_\Gamma^\trans \Phi_\Lambda} \le \norm{l2}{\vec{I} - \Phi^\trans_\Omega \Phi_\Omega} \le \ds
\end{equation*}
as the eigenvalues of $\vec{I} - \Phi^\trans_\Omega \Phi_\Omega$ are bounded above by $\ds$.  Hence
\begin{subequations}
\begin{align*}
	\norm{l2}{\Phi_\Gamma^\trans \Phi_\Lambda \x_\Lambda} &\le \norm{l2}{\Phi_\Gamma^\trans \Phi_\Lambda} \cdot\norm{l2}{\x_\Lambda} \\
	&\le\ds \norm{l2}{\x_\Lambda},
\end{align*}
\end{subequations}
completing the lemma.
\end{proof}

We now have the necessary results to prove Lemma \ref{lem:main}.
\begin{proof}[Proof of Lemma \ref{lem:main}]  Note that this proof follows parts of the proof in \cite{Blumensath2008}. 
First put
\begin{subequations}
\begin{align*}
	\r^{(i)} &\define \x - \x^{(i)}, \\
	\a^{(i)} &\define \x^{(i-1)} + \gamma\Phi^\trans(\y - \Phi\x^{(i-1)}), \\
	\x^{(i)} &\define \Ht_s(\a^{(i)}), \\
	\Gamma^\star &\define \supp(\x), \\
	\Gamma^{(i)} &\define \supp(\x^{(i)}), \\
	\it{B}{i} &\define \Gamma^\star \cup \Gamma^{(i)}.
\end{align*}
\end{subequations}
As a consequence we have $\abs{\Gamma^\star} \le s$ and $\abs{\it{\Gamma}{i}} \le \sh$. 

Consider the error $\norm{l2}{\x^s - \x^{(i+1)}}$.  Now we have 
\begin{subequations}
\begin{align*}
	\x&=\x^s = \x_{\Gamma^\star} = \x_{\it{B}{i}} &&\text{and}\\
	\it{\x}{i} &= \it{\x}{i}_{\it{\Gamma}{i}} = \it{\x}{i}_{\it{B}{i}}.
\end{align*}
\end{subequations}
Applying the triangle inequality we get
\begin{equation} \label{eq:mfr_eq1}
	\norm{l2}{\x^s - \x^{(i+1)}} \le \norm{l2}{\x^s_{B^{(i+1)}} - \a^{(i+1)}_{B^{(i+1)}}} + \norm{l2}{\x^{(i+1)}_{B^{(i+1)}} -  \a^{(i+1)}_{B^{(i+1)}}}.
\end{equation}
As $\x^{(i+1)}$ is the thresholded version of $\a^{(i+1)}$ it is the best $s$-term approximation to $\a^{(i+1)}$, in particular it is better than $\x^s$.  Hence
\begin{equation*}
	\norm{l2}{\x^{(i+1)} - \a^{(i+1)}_{B^{(i+1)}}} \le \norm{l2}{\x^s - \a^{(i+1)}_{B^{(i+1)}}},
\end{equation*}
and thus \eqref{eq:mfr_eq1} becomes
\begin{equation*}
	\norm{l2}{\x^s - \x^{(i+1)}} \le 2 \norm{l2}{\x^s_{B^{(i+1)}} - \a^{(i+1)}_{B^{(i+1)}}}.
\end{equation*}
Using the fact that $\y = \Phi\x^s + \e$ and $\r^{(i)} = \x^s - \x^{(i)}$ we get
\begin{subequations}
\begin{align*}
	\a^{(i+1)}_{B^{(i+1)}} &= \x^{(i)}_{B^{(i+1)}} + \gamma \Phi^\trans_{B^{(i+1)}}\left(\y-\Phi\x^{(i)}_{B^{(i+1)}}\right) \\
	&=\x^{(i)}_{B^{(i+1)}} + \gamma \Phi^\trans_{B^{(i+1)}}\Phi\r^{(i)} + \gamma\Phi^\trans_{B^{(i+1)}}\e,
\end{align*}
\end{subequations}
hence
\begin{align}
	\norm{l2}{\x^s - \x^{(i+1)}} & \le 2 \Bigg\|\underbrace{\x^s_{B^{(i+1)}}-\x^{(i)}_{B^{(i+1)}}}_{\r^{(i)}_{B^{(i+1)}}}-\gamma\Phi^\trans_{B^{(i+1)}}\Phi\r^{(i)} -\gamma\Phi^\trans_{B^{(i+1)}}\e  \Bigg\|_2 \nonumber\\
	&\le 2\norm{l2}{\r^{(i)}_{B^{(i+1)}}-\gamma\Phi^\trans_{B^{(i+1)}}\Phi\r^{(i)}} + 2\gamma\norm{l2}{\Phi^\trans_{B^{(i+1)}}\e} \nonumber\\
	&= 2\left\| \left(\vec{I}-\gamma\Phi^\trans_{B^{(i+1)}}\Phi_{B^{(i+1)}}\right)\r^{(i)}_{B^{(i+1)}} - \right.\ldots \nonumber\\
	&\quad  \left.-\gamma\Phi^\trans_{B^{(i+1)}}\Phi_{B^{(i)}\setminus B^{(i+1)}} \r^{(i)}_{B^{(i)}\setminus B^{(i+1)}} \right\|_2 +2\gamma\norm{l2}{\Phi^\trans_{B^{(i+1)}}\e} \nonumber\\
	&\le 2\norm{l2}{\left(\vec{I}-\gamma\Phi^\trans_{B^{(i+1)}}\Phi_{B^{(i+1)}}\right)\r^{(i)}_{B^{(i+1)}}} + \ldots\nonumber \\
	&\quad +2\gamma\norm{l2}{\Phi^\trans_{B^{(i+1)}}\Phi_{B^{(i)}\setminus B^{(i+1)}} \r^{(i)}_{B^{(i)}\setminus B^{(i+1)}} } +2\gamma\norm{l2}{\Phi^\trans_{B^{(i+1)}}\e}, \label{eq:mfr_residual}
\end{align}
by repeated application of the triangle inequality and by splitting the residual into two parts, $\it{\r}{i+1} = \it{\r}{i+1}_{\it{B}{i+1}} + \it{\r}{i+1}_{\it{B}{i+1}\setminus\it{B}{i}}$.  Then 
\[
	\abs{B^{(i)}\cup B^{(i+1)}} = \abs{\Gamma^\star \cup \Gamma^{(i)} \cup \Gamma^{(i+1)}} \le s+2\sh
\]
as each set $\Gamma^{(k)}$ has at most $\sh$ entries and $\abs{\Gamma^\star}\le s$.  Recall from the RIP and Lemmas \ref{lem:blu_1} and \ref{lem:eig_values} that
\begin{subequations}
\begin{align}
	\norm{l2}{\Phi^\trans_\Lambda \x}^2 & \le (1+\ds)\norm{l2}{\x}^2, \label{eq:mfr_rip}\\
	\norm{l2}{\left(\vec{I} - \gamma \Phi^\trans_\Lambda\Phi_\Lambda\right)\x_\Lambda} & \le (1-\gamma(1-\ds))\norm{l2}{\x_\Lambda}, \label{eq:mfr_lem1}\\
	\norm{l2}{\Phi^\trans_\Omega\Phi_{\Omega'}\x_{\Omega'}} & \le \ds \norm{l2}{\x_{\Omega}}, \label{eq:mfr_lem2}
\end{align}
\end{subequations}
for all matrices $\Phi$ which obey the RIP and sets $\Lambda,\Omega,\Omega'$, where $\Omega,\Omega'$ are disjoint, $\abs{\Lambda}=s$ and $\abs{\Omega\cup\Omega'}=s$.  We also have $\delta_{s}\le\delta_{s'}$ for all positive integers $s\le s'$.  Applying \eqref{eq:mfr_lem1} to the first term in  \eqref{eq:mfr_residual}, and applying \eqref{eq:mfr_lem2} and \eqref{eq:mfr_rip} to the second and third terms respectively, we get
\begin{align}
	\norm{l2}{\r^{(i+1)}} &\le 2\big(1-\gamma \left(1-\delta_{s+\sh}\right)\big) \norm{l2}{\r^{(i)}_{B^{(i+1)}}} + \ldots \nonumber \\
	&\quad +2\gamma\delta_{s+2\sh}\norm{l2}{\r^{(i)}_{B^{(i)}\setminus B^{(i+1)}}} + \ldots \nonumber \\
	&\quad +2\gamma \sqrt{1+\dsh}\norm{l2}{\e}. \label{eq:mfr_res_case} 
\end{align}
The vectors $\r^{(i)}_{B^{(i+1)}}$ and $\r^{(i)}_{B^{(i)}\setminus B^{(i+1)}}$ are orthogonal as they have disjoint supports.  Now let $\u,\v\in\R^n$ be two orthogonal vectors, then
\begin{equation*}
	\norm{l2}{\vec{u}} + \norm{l2}{\vec{v}} \le \sqrt{2} \norm{l2}{\vec{u}+\vec{v}}.
\end{equation*}
We use this to bound the sum of the two terms $\norm{l2}{\r^{(i)}_{B^{(i+1)}}}$ and $\norm{l2}{\r^{(i)}_{B^{(i)}\setminus B^{(i+1)}}}$.

We first ask, how do the terms $ 1-\gamma \left(1-\delta_{s+\sh}\right)$ and $\gamma\delta_{s+2\sh}$ compare, given that $\delta_{s+\sh} \le \delta_{s+2\sh}$?  We then have either
\begin{subequations}
\begin{align}
	1-\gamma + \gamma\dsh &\le \gamma\dshh \iff \gamma \ge \frac{1}{\dshh-\dsh+1}, && \mbox{or} \label{eq:mfr_a}\\
	1-\gamma + \gamma\dsh &> \gamma\dshh\iff \gamma < \frac{1}{\dshh-\dsh+1}. \label{eq:mfr_b}
\end{align}
\end{subequations}
\begin{itemize} \listspacing
\item \textbf{Case 1 -- Equation \eqref{eq:mfr_a}:}  Equation \eqref{eq:mfr_res_case} becomes
\begin{align*}
	\norm{l2}{\it{\r}{i+1}} &\le 2\sqrt{2} \gamma \dshh \norm{l2}{\it{\r}{i}} + 2\gamma\sqrt{1+\dsh} \norm{l2}{\e}.
\end{align*}
Then if
\begin{equation*}
	2\sqrt{2}\gamma\dshh \le \frac 1 2 \iff \gamma\dshh \le \frac{1}{\sqrt{32}},
\end{equation*}
we have
\begin{equation*}
	\norm{l2}{\it{\r}{i+1}} \le\frac 1 2 \norm{l2}{\it{\r}{i}} + 2\gamma\sqrt{1+\dsh} \norm{l2}{\e}.
\end{equation*}
Hence
\begin{equation*}
	\norm{l2}{\x-\it{\x}{k}} \le 2^{-k}\norm{l2}{\x} + 4\gamma\sqrt{1+\dsh} \norm{l2}{\e},
\end{equation*}
provided that
\begin{subequations}
\begin{align*}
	\gamma &\ge \frac{1}{\dshh-\dsh+1}, && \mbox{and} \\
	\gamma \dshh &\le \frac{1}{\sqrt{32}} \approx 0.177. 
\end{align*}
\end{subequations}

\item \textbf{Case 2 -- Equation \eqref{eq:mfr_b}:}  Equation \eqref{eq:mfr_res_case} becomes
\begin{align*}
	\norm{l2}{\it{\r}{i+1}} &\le 2\sqrt{2} \big(1-\gamma(1-\bsh)\big) \norm{l2}{\it{\r}{i}} +2\gamma\sqrt{1+\dsh} \norm{l2}{\e}.
\end{align*}
If 
\begin{subequations}
\begin{align*}
	2\sqrt{2} (1-\gamma(1-\dsh)) \le \frac 1 2 
	&\iff \gamma(1-\dsh) \ge 1- \frac{1}{\sqrt{32}} \\
	&\iff \dsh \le 1-\frac{1}{\gamma} + \frac{1}{\gamma \sqrt{32}} = \frac{8\gamma - 8 +\sqrt{2}}{8\gamma},
\end{align*}
\end{subequations}
we again have
\begin{equation*}
	\norm{l2}{\it{\r}{i+1}} \le\frac 1 2 \norm{l2}{\it{\r}{i}} + 2\gamma\sqrt{1+\dsh} \norm{l2}{\e}.
\end{equation*}
Hence
\begin{equation*}
	\norm{l2}{\x-\it{\x}{k}} \le 2^{-k}\norm{l2}{\x} + 4\gamma\sqrt{1+\dsh} \norm{l2}{\e},
\end{equation*}
provided that
\begin{subequations}
\begin{align*}
	\gamma &< \frac{1}{\dshh-\dsh+1}, && \mbox{and} \\
	\gamma(1-\dsh) &\ge 1- \frac{1}{\sqrt{32}} \approx 0.82,
\end{align*}
\end{subequations}
\end{itemize}
Putting these two results together we have
\begin{equation*}
	\norm{l2}{\it{\r}{i+1}} \le\frac 1 2 \norm{l2}{\it{\r}{i}} + 2\gamma\sqrt{1+\dsh} \norm{l2}{\e},
\end{equation*}
if either of the following conditions (a) or (b) are met
\begin{subequations}
\begin{align}
	&\mbox{(a)} &\gamma \ge \frac{1}{\dshh-\dsh+1}, &\quad\mbox{ and }\quad \gamma \dshh \le \frac{1}{\sqrt{32}}, \quad \mbox{or} \label{eq:mfr_conda}\\
	&\mbox{(b)} &\gamma < \frac{1}{\dshh-\dsh+1}, &\quad\mbox{ and }\quad \gamma(1-\dsh) \ge 1- \frac{1}{\sqrt{32}}, \label{eq:mfr_condb}
\end{align}
\end{subequations}
completing the proof of the lemma.
\end{proof}

Looking at our algorithm in another way shows that the MFR algorithm is capable of attaining the bounds of Lemma \ref{lem:existence}.  Lemma \ref{lem:existence} says that given $\y=\Phi\x$, then the minimiser to $\norm{l0}{\xh}$ subject to $\y=\Phi\xh$ is unique and equal to $\x$ if $\delta_{2s}<1$.  Lemma \ref{lem:theoretical_max} and its corollary shows that the MFR algorithm can achieve this bound.
\begin{lem}\label{lem:theoretical_max}
Let $\x\in\Rmn$ be an $s$-sparse vector and let the matrix $\Phi\in\Rmn$ have RIP constants $\delta_s$ and assume we are given the measurements $\y=\Phi\x+\e$.  If 
\begin{equation}
	\frac{1-\frac{\alpha}{2}}{1-\dsh} < \gamma < \frac{1}{1-\dsh},
\end{equation}
for some constant $0<\alpha<1$ and $\sh\ge s$, and
\begin{equation} \label{eq:th_d}
	\dsh < 1,
\end{equation}
then the MFR algorithm produces an approximation $\it{\x}{k}$ satisfying
\begin{equation} 
	\norm{l2}{\x-\it{\x}{k}} \le \alpha^{k} \norm{l2}{\x} + \frac{2\gamma\sqrt{1+\dsh}}{1-\alpha} \norm{l2}{\e}.
\end{equation}
\end{lem}
\begin{proof}
Recall the fundamental step of the MFR algorithm
\begin{equation}
	\it{\x}{k+1} = \Ht_{\sh}\left(\it{\x}{k}+\gamma\Phi^\trans\left(\y-\Phi\it{\x}{k}\right)\right).
\end{equation}
As before, set $\it{B}{k}\define \supp(\x)\cup\supp(\it{\x}{k})$ and $\it{\a}{k+1} \define \it{\x}{k}+\gamma\Phi^\trans\left(\y-\Phi\it{\x}{k}\right)$. Then we have
\begin{subequations}
\begin{align*}
	\it{\a}{k+1}_{\it{B}{k+1}} &=  \it{\x}{k}_{\it{B}{k+1}}+\gamma\Phi_{\it{B}{k+1}}^\trans\left(\y-\Phi\it{\x}{k}_{\it{B}{k+1}}\right) \\
	&= \it{\x}{k}_{\it{B}{k+1}} + \gamma\Phi_{\it{B}{k+1}}^\trans\Phi(\x-\it{\x}{k}) + \gamma\Phi_{\it{B}{k+1}}^\trans\e.
\end{align*}
\end{subequations}
 This gives the error estimate
\begin{subequations}
\begin{align*}
	\norm{l2}{\x-\it{\x}{k+1}} 
	& \le \norm{l2}{\x-\it{\a}{k+1}_{\it{B}{k+1}}} + \norm{l2}{\it{\a}{k+1}_{\it{B}{k+1}}-\it{\x}{k+1}} \\
	&\le 2\norm{l2}{\x-\it{\a}{k+1}_{\it{B}{k+1}}} \\
	&=2\left\| \x-\it{\x}{k}_{\it{B}{k+1}}-\gamma\Phi_{\it{B}{k+1}}^\trans\Phi\left(\x-\it{\x}{k}_{\it{B}{k+1}}\right) - \gamma\Phi_{\it{B}{k+1}}^\trans\e \right\|_2 \\
	&\le 2\norm{l2}{\x-\it{\x}{k}_{\it{B}{k+1}}-\gamma\Phi_{\it{B}{k+1}}^\trans\Phi\left(\x-\it{\x}{k}_{\it{B}{k+1}}\right) }+ 2\norm{l2}{\gamma\Phi^\trans\e} \\
	&=2\norm{l2}{\left(\vec{I}-\gamma\Phi^\trans\Phi\right)\left(\x-\it{\x}{k}\right)} + 2\gamma\norm{l2}{\Phi_{\it{B}{k+1}}^\trans\e} \\
	&\le 2\left(1-\gamma\left(1-\dsh\right)\right) \norm{l2}{\x-\it{\x}{k}} +2\gamma\sqrt{1+\dsh}\norm{l2}{\e} ,
\end{align*}
\end{subequations}
by Lemmas \ref{lem:blu_1} and \ref{lem:eig_values}.  This implies that
\begin{align*}
	\norm{l2}{\x-\it{\x}{k}} &\le \left[2\Big(1-\gamma\left(1-\dsh\right)\Big)\right]^k \norm{l2}{\x} + \frac{2\gamma\sqrt{1+\dsh}}{1-2\Big(1-\gamma\left(1-\dsh\right)\Big)} \norm{l2}{\e},
\end{align*}
since $\it{\x}{0} = \vec{0}$.  Thus if $0<2\left(1-\gamma\left(1-\dsh\right)\right)\le\alpha < 1$ then the algorithm will converge.  Since
\begin{subequations}
\begin{align*}
	0< 2(1-\gamma(1-\dsh)) \le \alpha 
	&\iff 1-\frac{\alpha}{2} \le  \gamma(1-\dsh) < 1-\dsh \\
	&\iff \frac{1-\frac{\alpha}{2}}{1-\dsh} \le \gamma < \frac{1}{1-\dsh},
\end{align*}
\end{subequations}
the algorithm will converge provided $\dsh<1$ producing an approximation that obeys
\begin{equation*}
	\norm{l2}{\x-\it{\x}{k}} \le \alpha^k \norm{l2}{\x} + \frac{2\gamma\sqrt{1+\dsh}}{1-\alpha} \norm{l2}{\e},
\end{equation*}
completing the lemma.
\end{proof}
This lemma says that for the right value of $\gamma$ and provided that $\dsh<1$, the algorithm will always converge to the correct solution, at the cost of noise amplification.

\begin{cor}
Under the hypothesis of Lemma \ref{lem:theoretical_max} and if we could measure the signal $\y$ exactly, i.e.\ if $\e=\vec{0}$, then for some value of $\gamma$, setting $\sh=s$ gives an algorithm capable of attaining the bound in Lemma \ref{lem:existence}.
\end{cor}
\begin{proof}
Setting $\sh=s$ in \eqref{eq:th_d} gives the condition $\delta_{2s}<1$ and by choosing $\gamma$ so that
\begin{equation*}
	\frac{1-\frac{\alpha}{2}}{1-\dsh} \le \gamma < \frac{1}{1-\dsh},
\end{equation*}
then the algorithm will produce a sequence of approximations which converge to $\x$.
\end{proof}

What is interesting about these results, is that they rely only on the \emph{lower} RIP constant, that is, it only requires
\begin{equation*}
	0< 1-\delta_{2s} \le \frac{\norm{l2}{\Phi\x}^2}{\norm{l2}{\x}^2},
\end{equation*}
for all $2s$-sparse vectors $\x$.  This means that the MFR algorithm will recover the correct $s$-sparse solution $\x$ to $\y=\Phi\x$ provided that there are no $2s$-sparse vectors in the kernel of $\Phi$.  In fact this is a tight theoretical bound:\ assume that there exists a vector $\v\in\R^n$ that is $2s$-sparse and $\Phi\v=\vec{0}$.  Choose a set $\Gamma\subset\{1,2,\ldots,n\}$ of size $s$ so that $\Gamma\subset\supp(\v)$ and set $\x= -\v_\Gamma$ so that $\x$ is $s$-sparse.  Then
\begin{equation*}
	\y = \Phi\x = \Phi\x+\Phi\v = \Phi(\x+\v) = \Phi\u,
\end{equation*}
where $\u = \x+\v$ is  $s$-sparse and $\u\neq\x$.  Hence there is no unique minimiser to $\norm{l0}{\xh}$ subject to $\y=\Phi\xh$ and no algorithm will be able to return the correct solution 100\% of the time.  Thus the MFR algorithm is able to achieve the theoretically maximum performance.

Perhaps even more surprising is that the algorithm will converge as fast as we want (but still linearly), i.e.\ $\norm{l2}{\x-\it{\x}{k}}\le \alpha^k \norm{l2}{\x}$ for any $0<\alpha<1$, provided we can choose $\gamma$ so that $1-\frac{\alpha}{2} \le \gamma(1-\delta_{2s}) < 1$.  

This seems to be an astounding result, until we realise that this requires accurate values of $\delta_{2s}$ and explicitly calculating $\delta_{2s}$ for a random matrix is computationally equivalent to directly solving the $\ell_0$ minimisation problem $\arg\min\norm{l0}{\xh}$ subject to $\y=\Phi\xh$.

We introduce one final lemma before proving the main theorem. 

\begin{lem}[Reduction to sparse case, Lemma 6.1 of \cite{Needell2008}] \label{lem:sparse_reduction_case}
Let $\x$ be a vector from $\R^n$ and assume that $\Phi$ obeys the RIP of order $t$, then the sample vector $\y=\Phi\x+\vec{e}$ can also be written as $\y=\Phi\x^t + \tilde{\vec{e}}$ where
\begin{equation}
	\norm{l2}{\tilde{\vec{e}}} \le \sqrt{1+\delta_t} \left(\norm{l2}{\x-\x^t} + \frac{1}{\sqrt{t}} \norm{l1}{\x-\x^t}\right) + \norm{l2}{\vec{e}},
\end{equation}
for any $t\in\Z$.
\end{lem}

We now use Proposition \ref{prop:main} and Lemma \ref{lem:sparse_reduction_case} from \cite{Needell2008} to prove the main theorem.  This uses the same techniques as in \cite{Blumensath2008}.

 \begin{proof}[Proof of Theorem \ref{thm:main}]
 Let $\x^{s}$ be the best $s$-sparse approximation to $\x$.  Then observe that
 \begin{equation*}
 	\norm{l2}{\x-\it{\x}{k}} \le \norm{l2}{\x-\x^{s}} + \norm{l2}{\it{\x}{k} - \x^{s}}.
 \end{equation*}
 We then apply the algorithm to $\y$ to recover an $s$-sparse approximation.  From Proposition \ref{prop:main} we get the bound 
 \begin{equation*}
 	\norm{l2}{\it{\x}{k} - \x^{s}} \le 2^{-k} \norm{l2}{\x^{s}} + 4\gamma\sqrt{1+\delta_{2s}} \norm{l2}{\hat{\e}},
 \end{equation*}
 where $\hat{\e} = \y-\Phi\x^{s}$.  Using Lemma \ref{lem:sparse_reduction_case} and setting $t=2s$  we can write $\y=\Phi\x^{2s}+\tilde{\e}$ where $\x^{2s}$ is a best $2s$-sparse approximation to $\x$, such that
 \begin{equation*}
 	\norm{l2}{\tilde{\vec{e}}} \le \sqrt{1+\delta_{2s}} \left(\norm{l2}{\x-\x^{2s}} + \frac{1}{\sqrt{2s}} \norm{l1}{\x-\x^{2s}}\right) + \norm{l2}{\vec{e}},
 \end{equation*}
 Hence combining Lemma \ref{lem:sparse_reduction_case} and Proposition \ref{prop:main} we get
 \begin{align*}
 	\norm{l2}{\it{\x}{k} - \x^{s}} &\le 2^{-k} \norm{l2}{\x^{s}}+ 4\gamma\sqrt{1+\delta_{2s}}\norm{l2}{\e} + \ldots \nonumber \\
 	&\quad +4\gamma\big(1+\delta_{2s}\big)\left(\norm{l2}{\x-\x^{2s}} + \frac{1}{\sqrt{2s}}\norm{l1}{\x-\x^{2s}}\right),
 \end{align*}
 under conditions (a), (b) or (c).  This completes the proof of the theorem.
 \end{proof}

Note that in condition (a), namely \eqref{eq:mfr_conda}, setting $\gamma=1/(1+\ds)$ and $\sh=s$ gives the same conditions on convergence for the proof of the IHT algorithm in \cite{Blumensath2008}.  Our result however generalises and shows that decreasing the step length can compensate for larger $\delta$ values.  We choose to use the RIP rather than the modified-RIP, unlike the authors of \cite{Blumensath2008}, as it leads to easier implementation.  To implement IHT one requires a measurement matrix scaled by $1/(1+\ds)$, but it is unfeasible to perform this operation exactly.  Hence by avoiding this scaling and choosing a deliberately smaller $\gamma$ (which admittedly disguises some of the difficulty in this scenario) we can much more easily implement our algorithm.

Observe also that this theorem implies that as $\gamma \rightarrow 0$, the error due to the noise component in the model also goes to 0.

Strictly speaking, there is no real reason to require $\norm{l2}{\it{\r}{k+1}} \le \frac 1 2\norm{l2}{\it{\r}{k}} + 2\gamma\sqrt{1+\delta_{2s}}\norm{l2}{\e}$, any value $0\le \alpha < 1$ with $\norm{l2}{\it{\r}{k+1}} \le \alpha\norm{l2}{\it{\r}{k}} + 2\gamma\sqrt{1+\delta_{2s}}\norm{l2}{\e}$ would suffice, but perhaps offer significantly slower convergence.  What happens to the convergence conditions if we allow a larger $\alpha$?

Assume we have $\norm{l2}{\it{\r}{k+1}} \le \alpha\norm{l2}{\it{\r}{k}} + 2\gamma\sqrt{1+\delta_{2s}}\norm{l2}{\e}$, then it follows that
\begin{subequations}
\begin{align*}
	\norm{l2}{\it{\r}{k}} &\le \alpha \norm{l2}{\it{\r}{k-1}} + 2\gamma\sqrt{1+\delta_{2s}}\norm{l2}{\e} \\
	&\le \alpha^k \norm{l2}{\it{\r}{0}} + 2\gamma\sqrt{1+\delta_{2s}} \norm{l2}{\e} \sum_{i=1}^k \alpha^i \\
	&< \alpha^k \norm{l2}{\it{\r}{0}} + \frac{2\gamma\sqrt{1+\delta_{2s}}}{1-\alpha} \norm{l2}{\e},
\end{align*}
\end{subequations}
which unfortunately threatens significant noise amplification, especially as $\alpha$ gets close to 1.  Then the convergence criteria become
\begin{subequations}
\begin{align*}
	&\mbox{(a) } \gamma \ge \frac{1}{\dshh-\dsh+1},\quad \mbox{ and } 
	\quad \gamma \dshh \le \frac{\alpha}{\sqrt{8}} < \frac{1}{\sqrt{8}} \approx 0.35, \quad\mbox{ or} \\
	&\mbox{(b) } \gamma < \frac{1}{\dshh-\dsh+1}, \quad \mbox{ and } 
	\quad \gamma(1-\dsh) > 1- \frac{\alpha}{\sqrt{8}} > 1-\frac{1}{\sqrt{8}} \approx 0.65,
\end{align*}
\end{subequations}
which are slightly looser than before.

\subsection{Convergence}
If Algorithm \ref{alg:mfr_alg} converges, it either converges to the correct sparse solution, or it converges to another sparse vector, but one that is not an (approximate -- to some level of tolerance) solution to the equation $\y=\Phi\x+\e$.  If the algorithm converges to an incorrect vector $\xh$, it is simple to test this since $\norm{l2}{\Phi\xh-\y}\gg \norm{l2}{\e}$, and, if necessary, rerun the algorithm with different $\gamma$ or $\sh$ estimates.

Suppose that the algorithm converges to $\xh$ and that we know $\y$ exactly, i.e.\ we are investigating an equation of the form $\y=\Phi\x$.  Note that the solution returned by the algorithm, $\xh$, is $\sh$-sparse.  Let $\PP\in\R^{n\times n}$ be a diagonal matrix that is the projection matrix onto the components specified by the support of $\xh$, i.e.\ onto the non-zero components of $\xh$.  Observe that we have $\PP^2=\PP=\PP^\trans$ and $\PP\xh = \xh$.  Then the solution $\xh$ obeys
\begin{subequations}
\begin{align*}
	\xh &= \PP\left[ \xh + \gamma \Phi^\trans \Phi\left(\x-\xh\right)\right] \\
	\Rightarrow \PP \Phi^\trans \Phi \x &=  \PP \Phi^\trans \Phi \xh \\
	&= \PP^\trans \Phi^\trans \Phi \PP\xh.
\end{align*}
\end{subequations}
Recall that the Moore-Penrose pseudo-inverse $A^\dagger$ of a matrix $A$ is given by, if it exists, $A^\dagger = \left(A^*A\right)^{-1}A^*$.  Let $A = \Phi \PP$, then $\xh$ obeys
\begin{equation*}
	A^\trans\Phi\x = A^\trans A \xh.
\end{equation*}
If we can invert $A^\trans A$, then $\xh$ is given by
\begin{equation*}
	\xh = (A^\trans A)^{-1}A \Phi^\trans \x = \left(\Phi \PP\right)^{\dagger} \Phi \x.
\end{equation*}
This result is not at all surprising, it merely confirms that if we knew the support of $\x$, then we could find the solution by solving the least squares problem on this support.

\section{Modifications} \label{sec:modifications}
We propose several modifications to the MFR algorithm that increase both the success rate and the rate of convergence.

\subsection{Accelerated Polynomial Convergence}
We can easily apply the methods from Section \ref{sec:accelerated} to increase the convergence rate of our algorithm to get Algorithm \ref{alg:mfr_accelerated}.

\begin{algorithm}[!t] 
\begin{flushleft}
\caption{MFR with Polynomial Convergence} \label{alg:mfr_accelerated}
\Input 
\begin{itemize} 
	\item The measurement matrix $\Phi$.
	\item Observation vector $\y$.
	\item Estimate of sparsity $\hat{s}$ of the vector $\x$.
	\item Step size $\gamma$.
        \item Tolerance parameter $\etol$.
\end{itemize}

\Output
\begin{itemize} 
	\item A vector $\xh\in\R^n$ that is $\sh$-sparse.
\end{itemize}

\begin{algorithmic}[1]
  \State $\it{\omega}{0}\leftarrow 1$
  \State $\mu \leftarrow \frac{\sigma_{\max}^2(\Phi) - \sigma^2_{\min}(\Phi)}{\sigma_{\max}^2(\Phi) + \sigma^2_{\min}(\Phi)}$
  \State $\it{\x}{0}\leftarrow \vec{0}$
  \State $\it{\x}{1} \leftarrow \Phi^\trans \y$
  \State $k\leftarrow 1$
  \While{ $ \norm{l2}{\it{\x}{k} - \it{\x}{k-1} } \ge \etol$}
  \State $\it{\omega}{k+1} \leftarrow \frac{1}{1-\it{\omega}{k} \frac{\mu^2}{4}}$
  \State $\it{\x}{k+1} \leftarrow \Ht_{\hat{s}}\left[ \it{\x}{k-1} + \it{\omega}{k+1} \left( \gamma \Phi^\trans \left(\y-\Phi\it{\x}{k}\right) + \it{\x}{k} - \it{\x}{k-1} \right)\right]$ 
  \State $k\leftarrow k+1$
  \EndWhile
  \State \Return $\it{\x}{k}$
\end{algorithmic}
\end{flushleft}
\end{algorithm}

\subsection{Least Squares}
Another method to speed up convergence of the MFR algorithm, is to add a least squares minimisation step. The algorithm produces a sparse approximation to the solution, which we use as a method of selecting the columns of the solution that contain the non-zero data points.  On this set of columns, say $\Gamma$, we then solve the least squares problem
\begin{equation}
	\argmin_{\z\colon \supp(\z) = \Gamma} \norm{l2}{\y-\Phi\z},
\end{equation}
which has a convenient closed form solution.  As this algorithm is deterministic, solving this problem twice when restricted to the same support, gives the same solution.  Hence we will run the update step until the support of the largest $\sh$ components changes, and then solve the least squares problem restricted to this support.  The full algorithm is given as Algorithm \ref{alg:mfr_ls}.

\begin{algorithm}[t] 
\begin{flushleft}
\caption{MFR with Least Squares} 
\label{alg:mfr_ls}

\Input 
\begin{itemize}
	\item The measurement matrix $\Phi$.
	\item Observation vector $\y$.
	\item Estimate of sparsity $\hat{s}$ of the vector $\x$.
	\item Step size $\gamma$.
\end{itemize}

\Output
\begin{itemize} 
	\item A vector $\xh\in\R^n$ that is $\sh$-sparse.
\end{itemize}

\begin{algorithmic}[1]
  \State $\it{\x}{0}\leftarrow \vec{0}$
  \State $\it{\Gamma}{0} \leftarrow \emptyset$
          \State $k\leftarrow 1$
	\While{ $ \norm{l2}{\it{\x}{k} - \it{\x}{k-1} } \ge \etol$}
  \State $\xh \leftarrow \it{\x}{k}$ 
  \State $\xh \leftarrow  \xh + \gamma \Phi^\trans\left(\y - \Phi\xh\right)$ 
  \State $\xh\leftarrow \Ht_{\sh}(\xh)$ \Comment{Prune}
  \State $\Gamma \leftarrow \supp(\xh)$
  \If{$\Gamma \neq \it{\Gamma}{0} $}
  \State $\it{\x}{k+1} \leftarrow \Phi^\dagger_{\Gamma} \y$ \Comment{Solve LS on $\Gamma$}
  \State $\it{\Gamma}{0}\leftarrow \Gamma$
  \Else
  \State $\it{\x}{k+1} \leftarrow \xh$
  \EndIf
                \State $k\leftarrow k+1$
	\EndWhile
  \State \Return $\it{\x}{k}$
\end{algorithmic}
\end{flushleft}
\end{algorithm}

Furthermore, we can combine the polynomial variant of this algorithm with least squares.

\subsection{Adaptive Step-length}
So far we have only considered a step-length, $\gamma$, of fixed size, but inherently there is no reason why we cannot vary $\gamma$ from one iteration to the next.  Algorithms with a variable step-length are certainly well known, such as in soft-thresholding \cite{Daubechies2007} and gradient pursuit methods \cite{Blumensath2008d}.  

What we propose is a greedy strategy, so that at each iteration the step length $\it{\gamma}{k}$ is chosen so as to minimise a certain quantity.  One obvious choice is to minimise the $\ell_2$ norm of the residual, so that we would choose the step-length $\gk$ to minimise $\norm{l2}{\x-\it{\x}{k}}$, i.e.
\begin{equation*}
	\gk = \arg\min_{\gamma\ge 0} \norm{l2}{\x - \Ht_{\sh}\left(\it{\x}{k-1}+\gk\Phi^\trans\left(\y-\Phi\it{\x}{k}\right)\right)}.
\end{equation*}
Unfortunately, since we do not know $\x$, we cannot calculate this quantity.  An alternative then is to minimise the residual in $\Phi$-space, i.e.\ $\norm{l2}{\y-\Phi\it{\x}{k}}$, so at every iteration we choose $\gk$ so as to minimise
\begin{equation}
	\norm{l2}{\y - \Phi\Ht_{\sh}\left(\it{\x}{k-1}+\gk\Phi^\trans\left(\y-\Phi\it{\x}{k}\right)\right)}. \label{eq:est_gamma}
\end{equation}
This gives us Algorithm \ref{alg:mfr_gamma}.

\begin{algorithm}[!t] 
\begin{flushleft}
\caption{MFR Algorithm with locally optimal $\gamma$} \label{alg:mfr_gamma}

\Input 
\begin{itemize} \listspacing
     \item The measurement matrix $\Phi$.
     \item Observation vector $\y$.
     \item Estimate of sparsity $\hat{s}$ of the vector $\x$.
     \item Tolerance parameter $\etol$.
\end{itemize}

\Output
\begin{itemize} \listspacing
	\item A vector $\xh\in\R^n$ that is $\sh$-sparse.
\end{itemize}

\begin{algorithmic}[1]
	\State $\it{\x}{0}\leftarrow \vec{0}$
        \State $k\leftarrow 1$
	\While{ $ \norm{l2}{\it{\x}{k} - \it{\x}{k-1} } \ge \etol$}
		\State $\gk \leftarrow \arg\min_{\gamma} \norm{l2}{\y-\Phi\Ht_{\hat{s}}\left( \it{\x}{k} + \gamma \Phi^\trans\left(\y - \Phi\it{\x}{k}\right)\right)}$
		\State $\it{\x}{k+1} \leftarrow \Ht_{\hat{s}}\left( \it{\x}{k} + \gamma \Phi^\trans\left(\y - \Phi\it{\x}{k}\right)\right)$ 
	 \State $k\leftarrow k+1$
	\EndWhile
	\State \Return $\it{\x}{k}$
\end{algorithmic}
\end{flushleft}
\end{algorithm}

This algorithm has two properties, firstly the $\ell_2$-norm of the residual is a non-decreasing function, as shown by the following lemma.
\begin{lem} \label{lem:mfr_est_gamma}
Let $\Phi\in\Rmn$ be a measurement matrix that obeys the RIP of order $s$ and let $\x\in\R^n$ be the $s$-sparse signal we are trying to reconstruct given the measurements $\y = \Phi\x$.  Using the MFR algorithm with adaptive step-length, Algorithm \ref{alg:mfr_gamma} where the step-length, is chosen at every iteration so as to minimise the residual in $\Phi$-space, then the $\ell_2$-norm of the residual is a non-increasing function of the number of iterations.
\end{lem}
\begin{proof}
Clearly if $\it{\gamma}{k}=0$ then we have $\it{\x}{k} = \it{\x}{k-1}$ hence $\norm{l2}{\Phi\it{\r}{k}} = \norm{l2}{\Phi\it{\r}{k-1}}$.  Thus setting $\it{\gamma}{k}=0$ does not increase the norm of the residual, hence minimising this for $\gamma > 0$ can only further decrease $\norm{l2}{\Phi\it{\r}{k}}$.
\end{proof}

Although Lemma \ref{lem:mfr_est_gamma} does not guarantee convergence to a vector $\xh$, it promises convergence to a vector $\Phi\xh$ that lies on a sphere around the measurements $\y$.  

Secondly, provided that $\gk$ meets the conditions of Theorem \ref{thm:main} then convergence is at least as fast, as the theorem does not depend on the particular value of $\gamma$ used at every iteration.

\section{Simulation Results} \label{sec:results}
In this section we simulate a number of the algorithms we have discussed.  All simulations use measurement matrices with Gaussian entries and the data signals were generated by randomly choosing a support, and then generating Gaussian entries with variance 1 for the non-zero components.  All algorithms were terminated when the change in output from one iteration to the next was less than $10^{-7}$ using the $\ell_2$ norm.  We classed a simulation run as successful if it produced an output that had the correct support.

\subsection{Success Rate}
\begin{figure}[ht]
\centering
\includegraphics[width=5in]{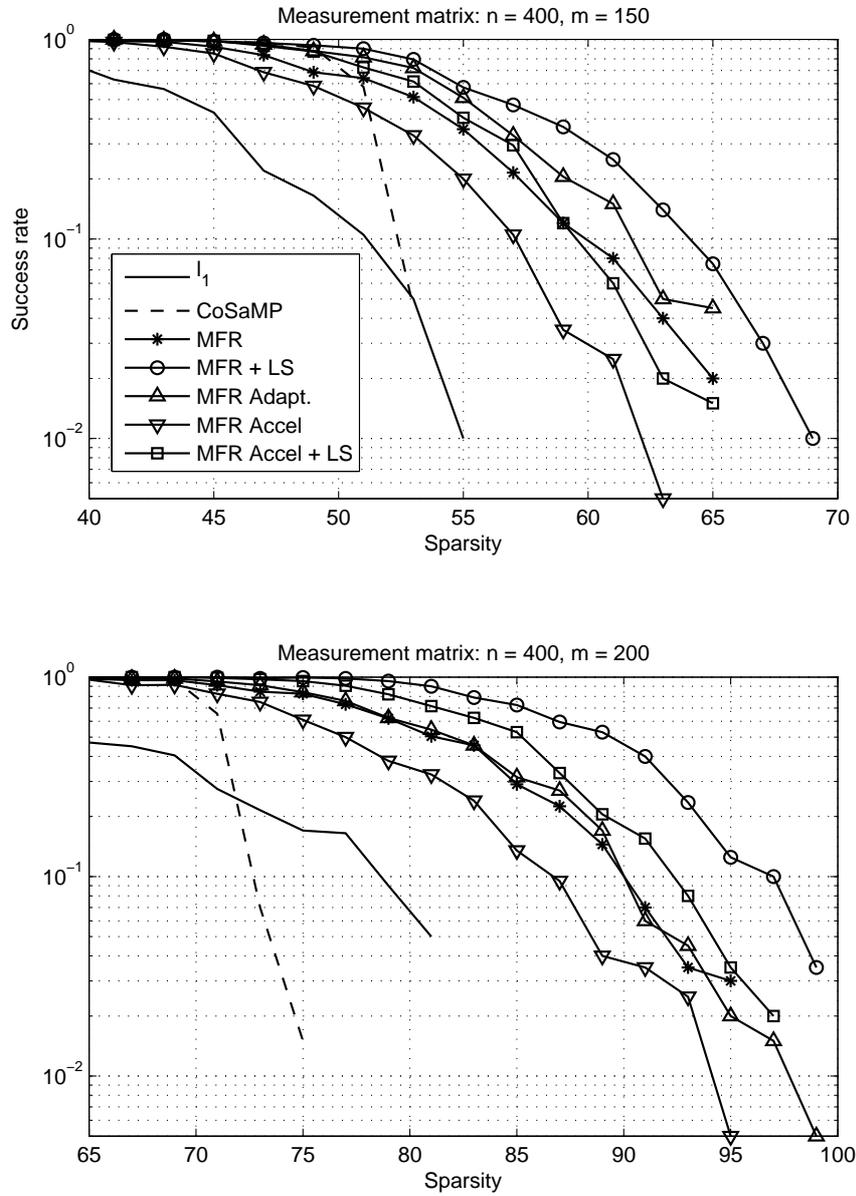}
\caption{Success rates for simulating a number of algorithms with $n=400$ and $m=150$ (top) and $m=200$ (bottom), using $\ell_1$ minimisation, CoSaMP, MFR and its variants. In both cases MFR with Least Squares outperforms all other algorithms.}
\label{fig:sim_results_n400}
\end{figure}

\begin{figure}[ht]
\centering
\includegraphics[width=5in]{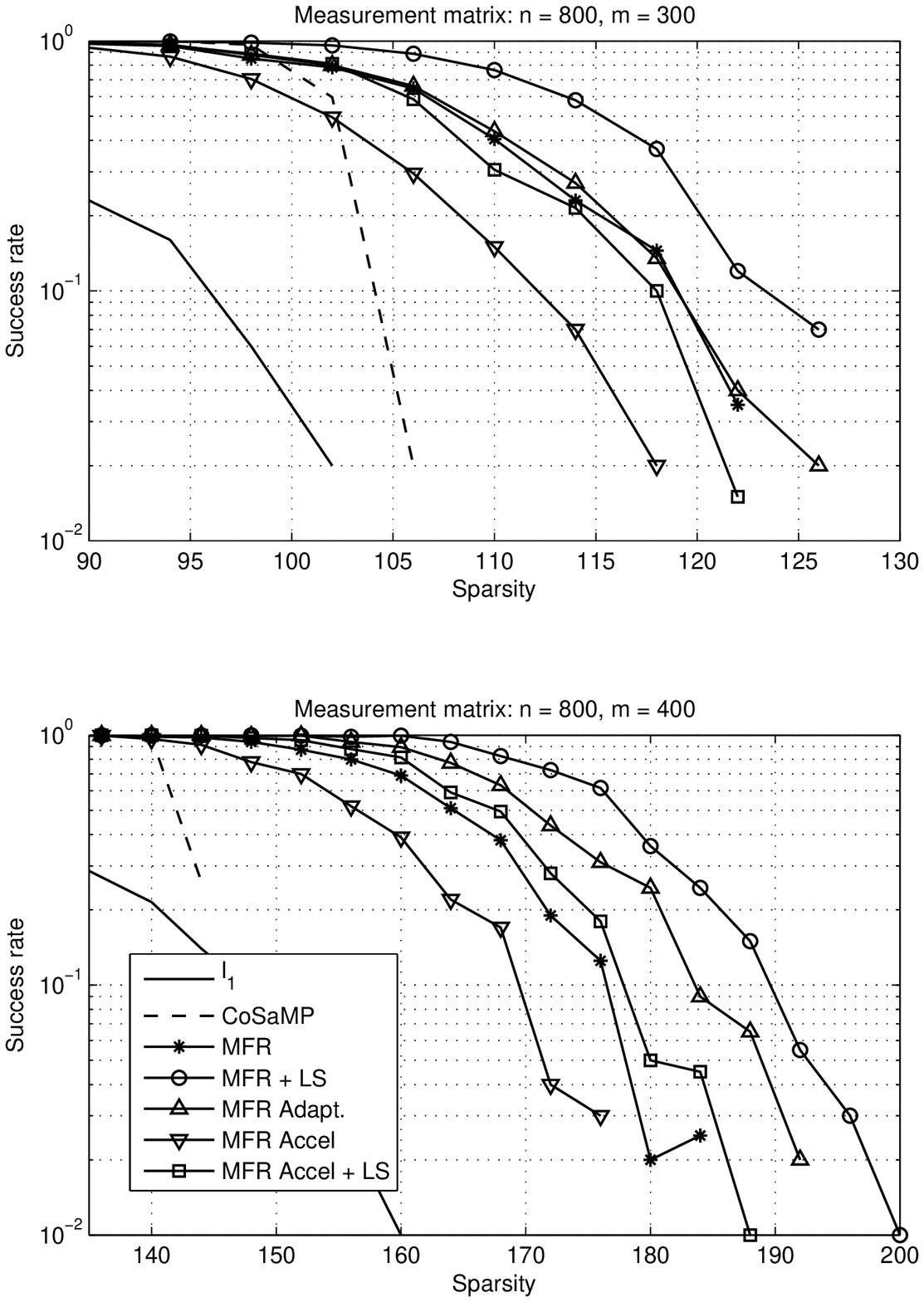}
\caption{Success rates for simulating a number of algorithms with $n=800$ and $m=300$ (top) and $m=400$ (bottom), using $\ell_1$ minimisation, CoSaMP, MFR and its variants.  In both cases MFR with Least Squares outperforms all other algorithms.}
\label{fig:sim_results_n800}
\end{figure}

In Figures \ref{fig:sim_results_n400}-\ref{fig:sim_results_n800} we plot the success percentages when simulating some of the algorithms mentioned in this paper.  In both cases we see that the MFR algorithm and its variants outperforms both $\ell_1$ minimisation and the CoSaMP algorithm.  The MFR algorithm with least squares (MFR + LS) is the best performing algorithm.  However what this graph does not capture is the speed and number of iterations required for convergence, which can be seen in Figure \ref{fig:mfr_convergence_rate}.

In Table \ref{tab:higher_se} we can see the advantage of choosing a sparsity estimate $\sh$ that is strictly larger than the true sparsity of the signal we are trying to reconstruct.  In every case, the single best sparsity estimate is one that is larger than the real sparsity.  This illustrates that it is in fact an advantage to overestimate the sparsity.  This is also significant as if we underestimate the sparsity, the algorithm cannot succeed.

\begin{table}[!t]
\caption{Percentage of simulations that resulted in the correct vector being returned by the MFR algorithm with least squares for a matrix $\Phi\in\R^{50\times 400}$.  Numbers in bold give the maximum for that column.  We see that we obtain a higher success rate if we underestimate the sparsity of the original signal.}
\label{tab:higher_se}
\centering
\begin{tabular}{|c||c|c|c|c|}
\hline
\textsc{Estimated} & \multicolumn{4}{c|}{\textsc{True Sparsity}} \\ 
\textsc{Sparsity} $\sh$        &  4   &  8  &  12 &   16 \\ \hline\hline
     4  &  26\%  &   -   &  -  &   - \\
     8  &  94\% &  11\%   &  -  &   - \\
    12  &  93\%  &  51\%   &  5\%  &   - \\
    16  &  \textbf{96\%}  &  \textbf{55\%}   &  7\%  &   0\% \\
    20  &  95\%  &  53\%   & \textbf{10\%}  &   \textbf{2\%} \\
    30  &  84\%  &  26\%   &  0\%  &   0\% \\
    40  &  41\%  &   6\%   &  0\%  &   0\% \\ \hline
\textsc{Total} & 100\% & 79\%& 17\% & 2\% \\ \hline
\end{tabular}

\end{table}

\subsection{Convergence Rate}
\begin{figure}[ht]
\centering
\includegraphics[width=4.4in]{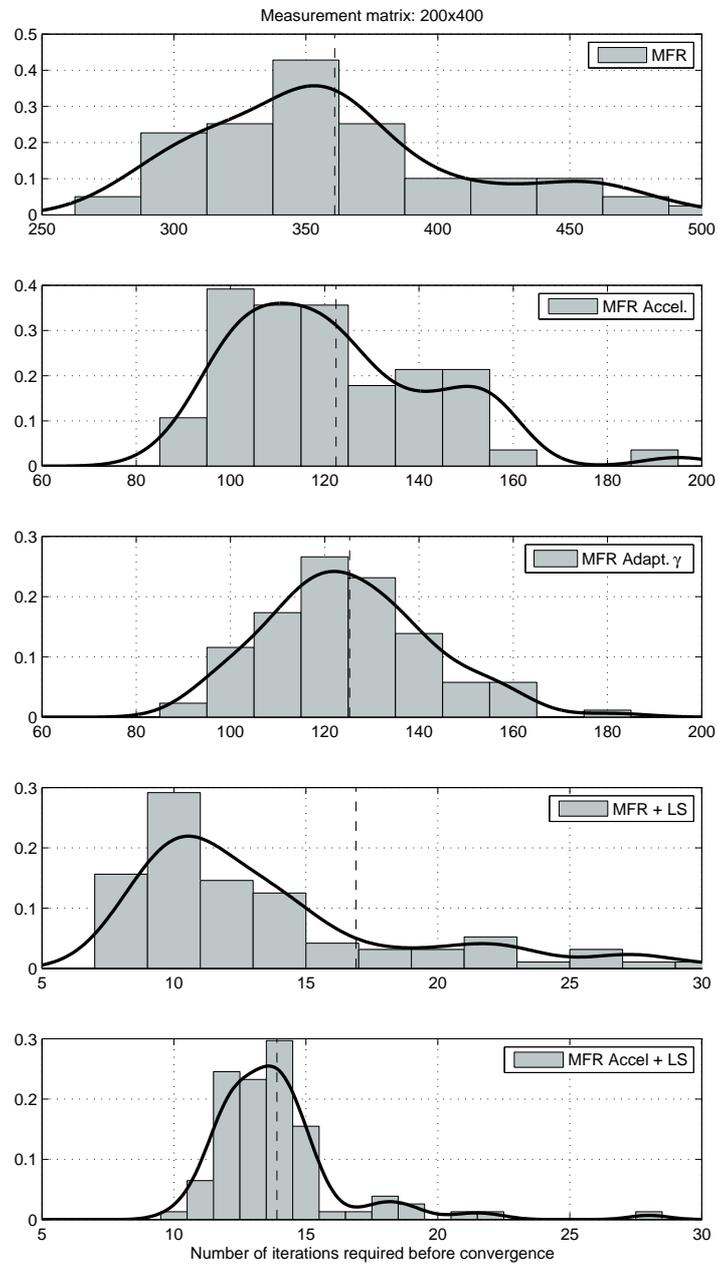}
\caption{Histogram of the number of iterations required for convergence for the MFR algorithms.  From top to bottom: MFR, MFR with polynomial acceleration, MFR with adaptive step-length, MFR with least squares and MFR with polynomial acceleration and least squares.  The vertical dashed line shows the mean number of iterations required before convergence.}
\label{fig:mfr_convergence_rate}
\end{figure}

In Figure \ref{fig:mfr_convergence_rate} we plot histograms of the number of iterations required before convergence for the three variants of our algorithms.  We generated one thousand random Gaussian matrices $\Phi$ and one thousand sparse data vectors $\x$, so that each algorithm was fed the same input.  We see clearly that the two modified versions converge significantly quicker than the plain MFR algorithm, and that the algorithm incorporating least squares converges faster still.

For a plot showing how the step-length affects the rate of convergence, see Figure \ref{fig:gamma_iterations}.

\clearpage

\section{Comparison to Previous Algorithms} \label{sec:comparison}
Here we present a detailed comparison of the MFR algorithm and its variants to several other reconstruction techniques.

The algorithm we have presented generalises the previous IHT algorithm.  We need to be careful in directly comparing the two algorithms as they use slightly different measurement matrix structures.  The IHT algorithm assumes a scaled matrix $\hat{\Phi}$, that is, let $\Phi\in\Rmn$ have RIP constant $\delta_s$ of order $s$, then the IHT algorithm reconstructs $\x$ given the measurements 
\begin{equation*}
	\y = \hat{\Phi}\x = \frac{\Phi}{1+\delta_s} \x,
\end{equation*}
whereas we operate directly with the original matrix $\Phi$.  This means that by setting $\gamma = \frac{1}{1+\delta_s}$, dividing the measurements $\y$ by $1+\ds$ and putting $\sh=s$ in our algorithm, the two algorithms are equivalent.

The advantage of allowing $\gamma$ to to be variable is phenomenal.  We have already shown that by allowing $\gamma$ to be smaller we can dramatically improve the reconstruction rate of the algorithm with low sparsity inputs.  Alternatively with very sparse inputs we can increase the convergence rate by choosing the value of $\gamma$ to be larger.  Proposition \ref{prop:main} says that, under the hypothesis of the theorem, the error in the final signal is bounded by the term $4\gamma \sqrt{1+\delta_{2s}}\norm{l2}{\e}$ where $\e$ is the error in measuring the signal.  Hence by taking a small value for $\gamma$ we can decrease the effect of the error in the final output.

\begin{figure}[!t]
\centering
\includegraphics[width=5in]{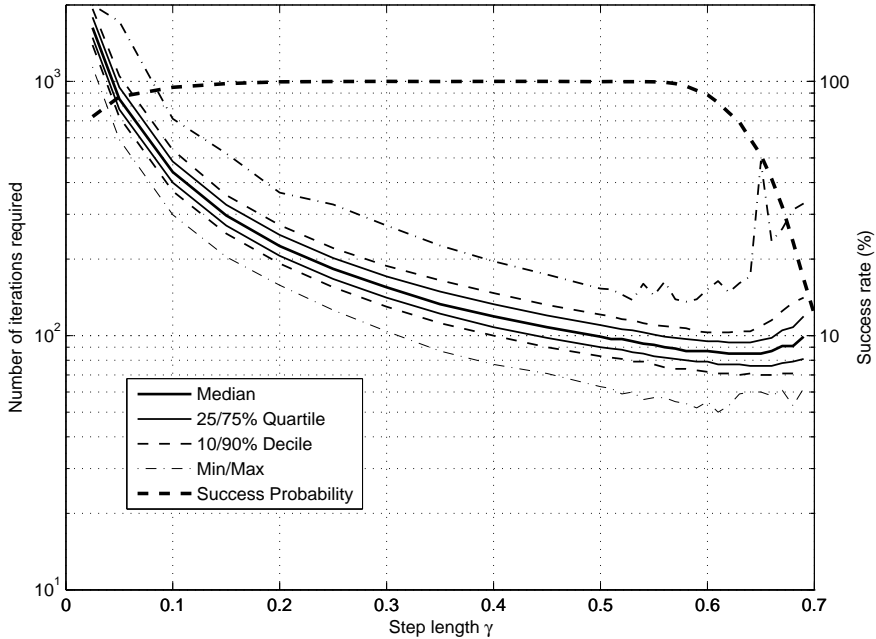}
\caption{Number of iterations required before convergence using the MFR algorithm and different (fixed) step-lengths.  This plot shows the quartiles for $\gamma\in\{0.05,0.1\ldots,0.7\}$.  Observe that as $\gamma$ increases the number of iterations required for convergence decreases, but the corresponding probability of success eventually falls.}
\label{fig:gamma_iterations}
\end{figure}

In comparison to the analysis of IHT, Theorem \ref{thm:main} offers a slightly better convergence result.  Setting $\sh=s$ and $\gamma=\frac{1}{1+\delta_s}$ yields the identical condition to the main Theorem of \cite{Blumensath2008}, our analysis  says that provided $\gamma \ge 1 / \left(\dshh-\dsh+1\right)$, if $\dshh > 1/\sqrt{32}$ then we can decrease $\gamma$ and still get convergence.  Alternatively, if $\dshh < 1/\sqrt{32}$ then we can increase $\gamma$ to get faster convergence, provided we could estimate the quantities $\dsh$ and $\dshh$ accurately.  The only way to do this currently, is to check all $\binom{n}{\tilde{s}}$ submatrices ($\tilde s = s+\sh$ or $\tilde s = s+2\sh$ as appropriate) of $\Phi$, which is computationally unfeasible and is in fact computationally equivalent to directly solving the original $\ell_0$ problem, $\arg\min \norm{l0}{\xh}$ subject to ${\y=\Phi\xh}$, directly.  We see in Figure \ref{fig:gamma_iterations} how the rate of convergence increases dramatically with $\gamma$, which is what we predicted from Theorem \ref{thm:main}.

By requiring a scaled measurement matrix, the implementation of the IHT algorithm becomes complicated.  There is no known way to generate measurement matrices with particular RIP constants, although we can say with high probability they are bounded by a certain value, but to implement the IHT algorithm requires the matrix to be scaled by $1+\delta_s$.  Hence we must either estimate $\delta_s$, which is not discussed in the original work, or calculate it explicitly, which means we might as well directly solve the $\ell_0$ problem.

We have also proposed adaptively choosing $\gk$ at every iteration.  We have shown that doing so can dramatically decrease the error rate in reconstruction.  It is ongoing work to see if there is a good way to estimate a better $\gamma$ in terms of minimising the residual, without directly knowing the residual.

Another important difference to the IHT algorithm is that we discuss what happens if we threshold with a value that is strictly larger than the sparsity of the signal we are trying to reconstruct, that is, we keep more than $s$ components of the signal non-zero.  We often see that choosing a larger thresholding value dramatically improves the success rate at the expense of increasing the number of iterations required for convergence.  

The algorithm we propose also appears on the surface to be very similar to the Gradient Pursuit algorithms discussed in \cite{Blumensath2008d}.   Indeed, the directional update in gradient pursuit is the same as for both IHT and MFR, but the big difference is in how the sparsity constraint is enforced.  For the gradient pursuit algorithms, a new dictionary element is added at every iteration, and once added, cannot be removed.  In contrast, IHT and MFR make use of a pruning step, so at every iteration we keep only the most important (decided by the largest magnitude) dictionary elements, thus elements can be both added and removed.

More recent work in \cite{Blumensath2009} uses the same paradigm as we do, i.e., selecting a possibly new support at each iteration, where the step-length is chosen to be the negative gradient of $\norm{l2}{\y-\Phi\it{\x}{k}}$, however their analysis is somewhat different to ours.

Incorporating a least squares step into the MFR algorithm makes it look similar to the CoSaMP algorithm in \cite{Needell2008}.  There are several significant differences though.

The CoSaMP algorithm solves the least squares problem over the vectors that have support size $3s$ where $s$ is the sparsity of the true solution.  The MFR algorithm typically considers a much smaller support size.  The column or support selection for the CoSaMP algorithm is performed by merging the support of the the previous iterate and the ``signal proxy'', that is $\Phi^\trans\Phi\it{\r}{k} = \Phi^\trans\Phi\left(\x - \it{\x}{k}\right)$.  In the case of the MFR algorithm, the frame reconstruction algorithm is used to select the support on which to solve the least squares problem.

Our work also addresses an area not explicitly discussed previously.  We frame our reconstruction results in terms of the true sparsity of the signal and an estimate of the sparsity.  Indeed we show via simulation that it can in fact be of benefit to under-estimate the sparsity of the signal we are trying to reconstruct.

\section{Conclusion}
In this paper we have demonstrated an iterative hard thresholding algorithm inspired by the frame reconstruction algorithm.  In this work we have generalised the results of \cite{Blumensath2008} and have shown how they can be used to guarantee convergence in a wider range of circumstances.  Furthermore our algorithm significantly outperforms the IHT algorithm at larger $s$ values (where $s$ is the sparsity of the signal we are tying to reconstruct).  At these points however, we operate beyond the range of values for which the theorem is applicable.  That is, the $\delta$ and $\gamma$ parameters do not satisfy either of our conditions because the sparsity level $s$ of $\x$ is too high.  However we see from out simulations that even though the algorithms are not guaranteed to converge to the correct solution, they still often do.

We  demonstrate several modifications to our algorithm;\ incorporating polynomial acceleration, adding a least squares step and using an adaptive algorithm to select the step-length.  These modifications increase the performance, that is the rate of successful reconstruction and the rate of convergence. 

For the least squares modification we show analytically that the algorithm will converge under the same requirements but possibly slightly slower, but empirically we observe that the algorithm in fact converges significantly faster.  For the adaptive step-length modification we show that provided the step-length satisfies the requirements of Theorem \ref{thm:main} the algorithm will converge at least as fast.

\bibliographystyle{plain}

\end{document}